# Intrinsic Spin Filter Effect in a *d*-wave altermagnet KV$_2$Se$_2$O with Open Fermi Surface

Bin Liu,[1] Pei-Hao Fu,[1,2,#] Yu-Xuan Sun,[1] Xiao-Lin Zhang,[3] Si-Cong Zhu,[4] Xiang-Long Yu, [1,*] Hua Wu, [5,†] Yuan-Zhi Shao [1,6,‡]

[1] School of Science, Sun Yat-sen University, Shenzhen 518107, China
[2] Department of Physics and Astronomy, University of Florence, I-50019 Sesto Fiorentino, Italy
[3] Key Laboratory of Artificial Micro- and Nano-structures of Ministry of Education, School of Physics and Technology, Wuhan University, Wuhan 430072, China
[4] The State Key Laboratory for Refractories and Metallurgy, Wuhan University of Science and Technology, Wuhan 430081, China
[5] Laboratory for Computational Physical Sciences (MOE), State Key Laboratory of Surface Physics, and Department of Physics, Fudan University, Shanghai 200433, China
[6] Guangdong Provincial Key Laboratory of Magnetoelectric Physics and Devices, Sun Yat-sen University, Guangzhou 510275, China

[#] *Corresponding author: phy.phfu@gmail.com*
[*] *Corresponding author: yuxlong6@mail.sysu.edu.cn*
[†] *Corresponding author: wuh@fudan.edu.cn*
[‡] *Corresponding author: stssyz@mail.sysu.edu.cn*

## ABSTRACT

Altermagnets offer a unique pathway to functional spintronics by combining vanishing magnetization with large spin splitting. Here, we demonstrate that the canonical *d*-wave altermagnet KV$_2$Se$_2$O can deliver giant tunneling magnetoresistance through orientation-dependent spin filtering. By analyzing the crystallographic spin segregation, we show that transport along specific crystallographic axes is nearly fully spin-polarized within the symmetry-protected ballistic channels. We implement this mechanism in a lattice-matched KV$_2$Se$_2$O/Bi$_2$O$_2$Se/KV$_2$Se$_2$O magnetic tunnel junction, which achieves a robust half-metallic transport regime. The symmetry-protected spectral gap in the parallel/anti-parallel configuration ensures a high tunneling magnetoresistance ratio, resulting in substantial tunneling magnetoresistance, robust thermally driven spin filtering, and spin Seebeck effect at room temperature. These findings provide a path of altermagnetic heterostructures as a high-performance platform for scalable, field-free, and thermally stable spin logic.

# 1. INTRODUCTION

Spin-polarized charge currents constitute the fundamental information carrier in spintronic technologies, underpinning key functionalities such as nonvolatile memory, logic operations, and spin-torque-driven switching[1]. Achieving robust spin polarization without sacrificing device scalability and stability remains a central challenge. Ferromagnets (FM) naturally generate longitudinal spin-polarized currents but suffer from stray magnetic fields that limit device density and introduce cross-talk[2-5]. Antiferromagnets (AFM) eliminate stray fields and exhibit ultrafast spin dynamics, yet their compensated spin symmetry typically enforces spin-degenerate electronic bands, resulting in negligible longitudinal spin polarization and severely restricting their applicability in spin-charge conversion and transport[6-9]. Altermagnetism (AM) has recently emerged as a distinct magnetic order that bridges this long-standing dichotomy by unifying key attributes of ferromagnets and antiferromagnets[10-12]. Altermagnets host large, exchange-driven spin splitting comparable to ferromagnets and unconventional spin Hall responses enabled by magnetic crystal symmetries, while maintaining a strictly vanishing net magnetization enforced by combined crystal rotation and time-reversal symmetries[13-17]. Owing to the absence of net magnetization, AM avoids the stray fields that challenge FM-based devices. At the same time, the presence of spin splitting enables spin-dependent transport, in contrast to AFM-based systems in which the spin information is largely hidden[18-23].

However, the symmetry protection that stabilizes altermagnetism also imposes a fundamental constraint: spin textures related by combined crystalline and time-reversal symmetries, exhibit an orientation-dependent compensation that leads to an exact cancellation of spin polarization when integrated over the Brillouin-zone, thereby prohibiting the emergence of sizable macroscopic spin-polarized charge currents in equilibrium. In this sense, the defining advantage of altermagnetism simultaneously constitutes its central obstacle for spintronic functionality[24-27]. From a symmetry perspective, both strain- and electric-field-based approaches rely on explicit symmetry breaking to enable spin polarization in altermagnets. Uniaxial strain induces fully spin-polarized transport by breaking mirror and rotational symmetries, analogous to electric-field-based schemes that rely on strong gate fields to lift symmetry constraints. Crucially, both mechanisms yield momentum-dependent and valley-confined spin responses[28, [29]. Such strong external perturbations raise concerns regarding robustness, scalability, and experimental feasibility. In terms of physical mechanisms, these strategies rely on explicit symmetry breaking to lift the symmetry-enforced momentum-space cancellation of spin polarization. The resulting spin-polarized transport can be interpreted as arising from an effective Zeeman-like field that induces a finite net magnetization[30-34]. While this route enables longitudinal spin polarization, it effectively drives the system back to a ferromagnetic/ferrimagnetic-like regime, undermining the symmetry-protected nature of altermagnetism and potentially reintroducing familiar limitations such as stray-field coupling and reduced device robustness. These considerations underscore a critical open challenge in altermagnetic spintronics: achieving intrinsic longitudinal spin-polarized charge transport without resorting to external symmetry-breaking fields or net magnetization.

In this work, we propose an intrinsic spin-filter effect in AM driven by open Fermi-surface geometry, enabling highly spin-polarized charge transport without net magnetization. Focusing on $KV_2Se_2O$ as a representative material platform, we first establish the microscopic mechanism underlying the intrinsic spin-filter effect using a simple low-energy effective model. We show that the closed-to-open Fermi surface transition selectively suppresses one spin channel in a direction-selective manner, resulting in ideal spin polarization in longitudinal transport. We then provide a proof of concept by investigating $KV_2Se_2O$—a material exhibiting strong crystalline and magnetic anisotropy that enables pronounced directional modulation of spin-dependent charge transport. Owing to its quasi-two-dimensional lattice, strong in-plane anisotropy, and intrinsic magnetic order, $KV_2Se_2O$ supports symmetry-constrained, direction-dependent spin textures, enabling field-free spin-polarized transport within a realistic materials platform[14, 35]. By combining first-principles electronic-structure calculations with device-level transport modeling, we elucidate the symmetry origin of the anisotropic spin splitting, track the evolution of momentum-resolved spin textures under different transport geometries, and identify specific crystallographic directions along which nearly approaching fully longitudinal spin polarization emerges despite the absence of net magnetization. Guided by the Density Functional Theory (DFT) results, we further propose lattice-matched tunneling heterostructures based on $KV_2Se_2O$. The sharp contrast between the conducting parallel-configuration (PC) state and the insulating antiparallel-configuration (APC) state yields a giant tunneling magnetoresistance (TMR) ratio of approximately $2.24 \times 10^3$ % in the absence of any external magnetic field. Moreover, we predict a colossal thermal tunneling magnetoresistance exceeding 750% at low temperatures, which remains substantial even at room temperature. Our findings demonstrate that altermagnets with open Fermi surfaces enable a distinct class of spintronic functionalities, including nearly fully spin filters and giant tunneling magnetoresistance, without requiring magnetic fields, spin--orbit coupling, or ferromagnetic components.

## 2. RESULTS

2.1 Low-Energy Effective Model for Fermi Surface Controlled Spin Transport

In this section, we use a low-energy effective model for a two-dimensional $d$-wave altermagnet to elucidate the mechanism of the geometry of the Fermi surface enabling theoretically perfect spin-polarized charge transport. The purpose of this effective model is to capture, in a minimal but analytically transparent manner, how the closed-to-open Fermi-surface transition in altermagnets controls (i) the density of states (DOS) and (ii) the direction-dependent spin-resolved carrier velocity, which together dictate the spin polarization of the longitudinal conductance.

The effective Hamiltonian for two-dimensional $d_{x^2-y^2}$-wave altermagnet is[10, 14]

$$\hat{H}_{AM}^{\sigma}(k_x, k_y) = t_x^{\sigma} a^2 k_x^2 + t_y^{\sigma} a^2 k_y^2 + E_F, \qquad (1)$$

where $a$ is the lattice constant and $E_F$ denotes the Fermi level, $\sigma = +/-$ for spin-up/down states. The $d$-wave altermagnetic anisotropic spin splitting is encoded in the spin- and direction-dependent hopping parameters $t_{x(y)}^{\sigma} = t_0 \pm \sigma t_{AM}$, with $t_0$ representing the isotropic nearest-neighbor hopping amplitude and $t_{AM}$ the anisotropic spin-dependent hopping that captures the $d$-wave altermagnetic spin splitting. This anisotropic spin splitting is directly manifested in the spin-resolved equal-energy contours and dispersions of Eq. (1), shown in Fig. 1. For $t_{AM}/t_0 < 1$ [Fig. 1(a)], the dispersion consists of two upward-opening parabolic bands with spin- and direction-dependent curvature, leading to two spin-dependent closed elliptical Fermi surfaces whose principal axes are mutually orthogonal [see the insert of Fig. 1(a)]. For $t_{AM}/t_0 > 1$ [Fig. 1(b)], the dispersion consists of two parabolic bands with opposite opening directions (along one momentum axis) and spin-dependent curvature. This yields open Fermi surfaces: within a finite momentum window, only a single spin species is present at the Fermi level as exhibited in the insert of Fig. 1(b). Moreover, the band structure hosts a saddle point at the spin-degenerate momentum $(k_x, k_y) = (0, 0)$[36], where the curvature is positive along one direction and negative along the other. Notably, at the critical point $t_{AM}/t_0 = 1$ [Fig. S1], a flat band emerges: along one direction, only one spin sub-band remains dispersive, while the other is dispersionless.

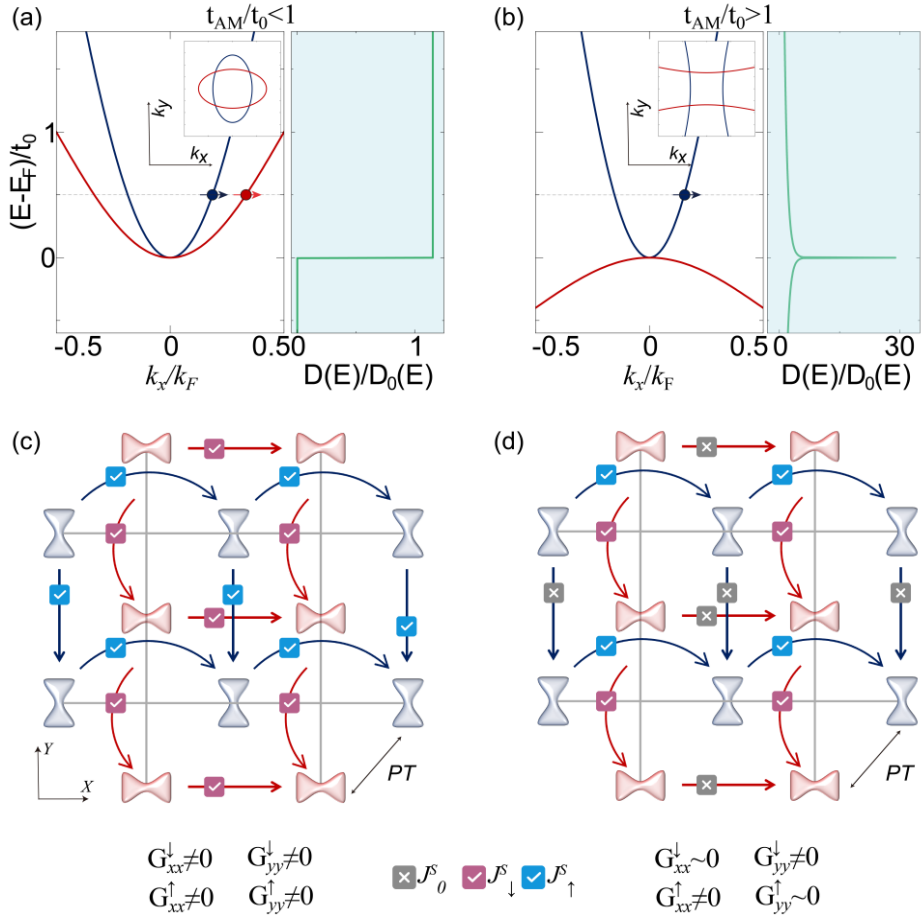

FIG. 1. Mechanism of open-Fermi-surface-driven perfectly spin-polarized transport. (a), and (b) Energy dispersion of the low-energy effective *d*-wave altermagnetic model [Eq. (1)] along $k_x$ with $k_y = 0$, together with the corresponding density of states [Eq. (2)] as a function of energy. Blue and red represent different spins, and the direction of the arrows indicates the direction of the speed. Insets illustrate the Fermi-surface geometry determined by the altermagnetic parameter $t_{AM}/t_0$: closed Fermi surfaces with positive curvature for $t_{AM}/t_0 < 1$, open Fermi surfaces with negative curvature for $t_{AM}/t_0 > 1$. (c), and (d) Schematic illustration of the spin-resolved transport corresponding to the different Fermi-surface geometries. For Fermi surfaces with nonpositive curvature ($t_{AM}/t_0 \geq 1$), direction-dependent and nearly perfectly spin-polarized charge transport emerges.

In the altermagnet with the open Fermi surface, the occurrence of the flat band and saddle point gives rise to a van Hove singularity in the DOS near the corresponding energy[36]. A straightforward calculation yields the spin-resolved DOS[37]

$$D^\sigma(E) = \frac{1}{8\pi} \frac{1}{\sqrt{|t_{AM}^2 - t_0^2|}} \begin{cases} 2\Theta(E - E_F), & \dfrac{t_{AM}}{t_0} < 1, \\ \ln\left(\dfrac{E_{cut}}{|E - E_F|}\right), & \dfrac{t_{AM}}{t_0} \geq 1, \end{cases} \quad (2)$$

where $E_{cut} = 8t_0$ is a bandwidth cutoff, $\Theta(x)$ is the Heaviside step function. Importantly, $D^\sigma(E) = D^{-\sigma}(E)$, so the spin density $D^\uparrow(E) - D^\downarrow(E) = 0$, consistent with the absence of net magnetization. The total DOS is then $D(E) = D^\uparrow(E) + D^\downarrow(E) = 2D^\sigma(E)$. Figs. 1(a-b) summarize the energy dependence of $D(E)$ for representative $t_{AM}/t_0$. For $t_{AM}/t_0 < 1$, $D(E)$ is essentially the constant DOS of a two-dimensional parabolic band[37], with modification induced by the altermagnetism via $t_{AM}$, but no singular behavior. By contrast, when $t_{AM}/t_0 \geq 1$, the flat band and saddle point generate a logarithmic van Hove singularity near $E \simeq E_F$. Notably, in the vicinity of the critical point $t_{AM}/t_0 = 1$, $D(E) \to \infty$ within the present effective description, implying a strong enhancement of the transport response. This sharp contrast between the DOS of closed and open Fermi surfaces is a direct signature of the distinct Fermi-surface geometries.

The impact of Fermi-surface geometry is also encoded in the carrier velocity. For Eq. (1), the velocity expectation value along $\alpha$-direction ($\alpha = x, y$) reads

$$v_\alpha^\sigma = \frac{1}{\hbar} \left\langle \partial_{k_\alpha} \hat{H}_{AM}^\sigma \right\rangle = \frac{2 t_\alpha^\sigma a^2 k_\alpha}{\hbar}, \quad (3)$$

where $v_\alpha^\sigma$ denotes the quantum-mechanical expectation value in the spin-$\sigma$ eigenstate. The $d$-wave altermagnetic symmetry implies $v_x^\sigma = v_y^{-\sigma}$, inherited from Eq. (1). As illustrated in Fig. 1(c), for $t_{AM} < t_0$ (closed Fermi surfaces), both spin channels have finite velocities along both directions, so spin-up and spin-down electrons contribute comparably to transport. In contrast, for $t_{AM} \gtrsim t_0$ [Figs. 1(d)], one spin channel becomes nearly blocked along a given direction while the other channel remains highly conductive. In particular, $v_x^\uparrow = v_y^\downarrow \propto t_0 + t_{AM}$, whereas $v_x^\downarrow = v_y^\uparrow \propto t_0 - t_{AM} \simeq 0$, so transport along $x(y)$ is dominated by spin-($\uparrow\downarrow$) carriers, providing the kinematic basis for perfectly spin-polarized currents.

The effect of the Fermi-surface geometry on the DOS [Eq. (2)] and the velocity [Eq. (3)] is then manifested in the longitudinal conductance. To demonstrate how the Fermi-surface geometry controls transport, we evaluate the longitudinal conductance within semiclassical Boltzmann theory using the relaxation-time approximation. In the linear-response regime, $I_\alpha = G_{\alpha\beta}V_\beta$, and we focus on the longitudinal component $\alpha = \beta$. At zero temperature, the conductance can be written as

$$G_{\alpha\alpha}(E) = \tau e^2 \sum_{\mathbf{k}} v_\alpha^2 D(E), \tag{4}$$

where $\tau$ is the momentum relaxation time. For the $x$ direction, substituting Eq. (3) into Eq. (4) yields the scaling form

$$G_{xx}^\sigma(E) \propto (v_x^\sigma)^2 D^\sigma(E) \sim t_0^2 \left(1 + \sigma \frac{t_{AM}}{t_0}\right)^2 D^\sigma(E), \tag{5}$$

which explicitly contains $t_{AM}/t_0$ through the spin- and direction-dependent velocity. For $|t_{AM}/t_0| < 1$, both $G_{xx}^\uparrow$ and $G_{xx}^\downarrow$ are comparable, and the net conductance contains contributions from both spins. For $|t_{AM}/t_0| \gtrsim 1$, however, $G_{xx}^\uparrow$ is strongly enhanced (particularly near the van Hove singularity at $E \simeq E_F$), whereas $G_{xx}^\downarrow$ becomes negligible, yielding a nearly perfect spin polarization

$$P_{xx} = \frac{G_{xx}^\uparrow - G_{xx}^\downarrow}{G_{xx}^\uparrow + G_{xx}^\downarrow} \simeq 1, \tag{6}$$

Repeating the same analysis for transport along $y$ gives the opposite polarization, $P_{yy} \simeq -1 = -P_{xx}$, consistent with the $d$-wave altermagnetic symmetry and the relation $G_{yy}^\sigma(E) = G_{xx}^{-\sigma}(E)$, as illustrated in Fig. 1.

We emphasize that Eq. (1) is a minimal effective description. In realistic materials such as $RuO_2$[38-41], $La_2O_3Mn_2Se_2$, and $Ba_2CaOsO_6$[42], $d$-wave altermagnetism often involves additional orbital and sublattice degrees of freedom (e.g., two-sublattice tetragonal[38-40] or Lieb-lattice models[42]). Here we isolate the universal role of Fermi-surface geometry, for which Eq. (1) already captures the key distinction between closed and open Fermi surfaces. We also note that the divergence in Eq. (2) is regularized in real systems by disorder and finite quasiparticle lifetimes[43]. A standard phenomenological broadening amounts to the replacement $|E - E_F| \mapsto \sqrt{(E - E_F)^2 + \Gamma^2}$, with $\Gamma \sim \hbar/(2\tau')$ set by the elastic lifetime $\tau'$ due to impurity scattering[44-46]. Finally, the Boltzmann treatment in Eq. (4) assumes a constant (isotropic) relaxation time $\tau$, weak disorder, zero temperature, and a small bias $V_\beta$. Despite these approximations, the analysis demonstrates the central point: open Fermi surfaces naturally suppress one spin channel in a direction-dependent manner, enabling giant and nearly perfectly spin-polarized longitudinal conductance when the chemical potential is tuned close to the flat-band or saddle-point energies. In the next section, we substantiate this mechanism using material-based calculations.

2.2 Nearly Perfectly Directional Spin Polarization in $KV_2Se_2O$

To validate the mechanism proposed in Eq. (1), in this section and the following we perform DFT calculations for the altermagnetic candidate material $KV_2Se_2O$. The calculations were performed using the nonequilibrium Green's function formalism combined with density functional theory (DFT+NEGF), as implemented in QuantumATK (version T-2022.03, Synopsys QuantumATK)[47-49], A Hubbard $U$ value of 1 eV was applied to the V $3d$ orbitals of one pair of V atoms with opposite spin orientations[50-52], thereby explicitly treating spin polarization. The electrically-driven spin transport performance can be quantitatively analyzed from the Landauer-Buttiker equation[53]:

$$I(V_b) = \frac{2e^2}{h} \int_{-\infty}^{+\infty} T(E, V_b)[f_L(E - \mu_L) - f_R(E - \mu_R)]dE \quad (7)$$

where $f_{L(R)}(E, \mu)$ is the equilibrium Fermi distribution of the left (right) electrode, and $\mu$ is the electrochemical potential of the electron. Fig. 2(a) shows the crystal structure of $KV_2Se_2O$, where the magnetic V atoms form a square lattice separated by Se and O ligands. The corresponding spin-resolved Fermi surfaces are shown in Fig. 2(b). Two bands (labeled as Bands 35 and 36) cross the Fermi energy and exhibit pronounced anisotropy: one spin sub band is dispersive along one direction (e.g. spin-up is dispersive along Γ-X direction, spin-down is dispersive along Γ-Y direction), it becomes flat band without dispersion along the orthogonal direction. As a result, the Fermi surfaces are open, extending across the Brillouin zone rather than forming closed contours. As what is expected in other altermagnetic candidate materials such as MnTe, $Nb_2FeB_2$, and $V_2Se_2O$[54-56]. This geometry closely resembles the open-Fermi-surface regime of the effective model Eq. (1) for $t_{AM} \gtrsim t_0$.

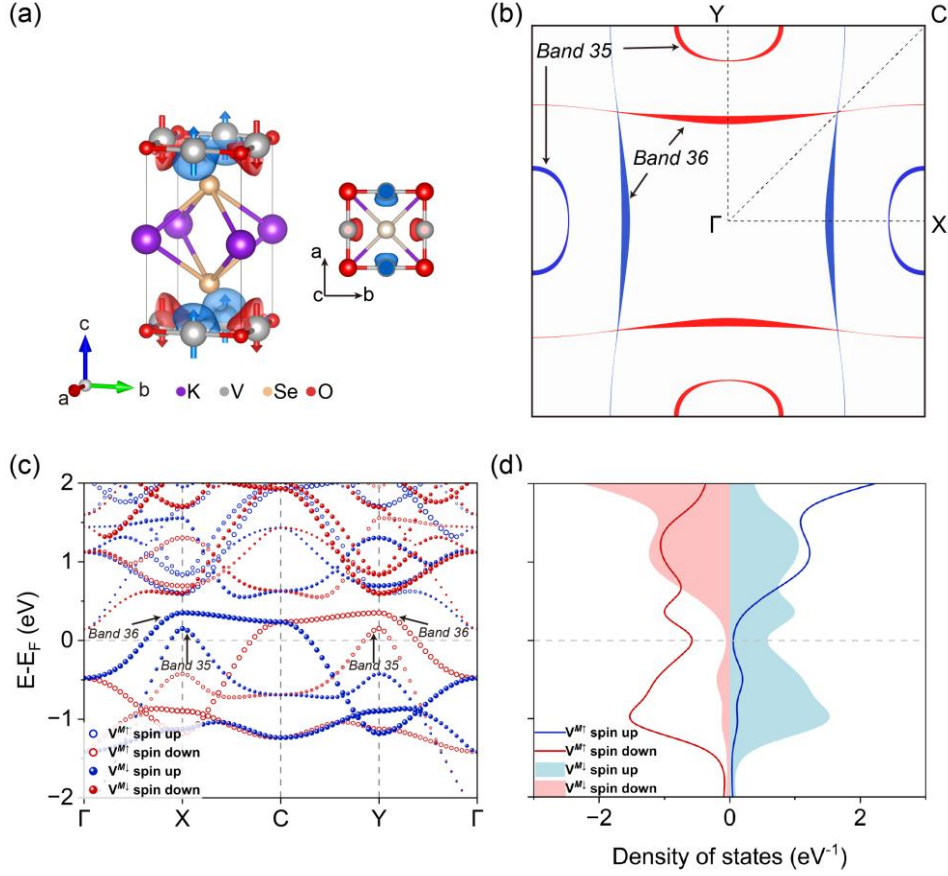

FIG. 2. Realization of crystallographic spin segregation in the *d*-wave altermagnet $KV_2Se_2O$. (a) Crystal and magnetic structure of $KV_2Se_2O$. Vanadium (V) atoms form two antiferromagnetically aligned magnetic sublattices ($V^{M\uparrow}$ and $V^{M\downarrow}$), which are related by a crystallographic rotation symmetry. (b) Spin-resolved Fermi surface in the $k_z = 0$ plane. The Fermi surface exhibits complete spin separation: electronic states along the Γ–X direction is exclusively spin-up (blue), whereas those along the orthogonal Γ–Y direction is entirely spin-down (red). (c) V-atom projected spin-resolved band structure. The marker size is proportional to the spectral weight from the $V^{M\uparrow}$ (open circles) and $V^{M\downarrow}$ (solid circles) sublattices, with blue and red denoting spin-up and spin-down states, respectively. Transport channels along Γ–X are dominated by $V^{M\uparrow}$ spin-up states, while those along Γ–Y are dominated by $V^{M\downarrow}$ spin-down states, demonstrating sublattice-selective transport. (d) Projected density of states (PDOS) of the V 3*d* orbitals. The spin-symmetric total spectral weight reflects complete magnetization compensation, while the pronounced local exchange splitting ensures robust metallic transport channels.

The origin of this open Fermi surface can be traced back to the orbital character and bonding geometry of $KV_2Se_2O$. As a results, in Fig. 2(c), the band structure near the Fermi level consists of two parabolic bands with opposite opening directions (along one momentum axis) and spin-dependent curvature, as shown in Fig. 1(b) via the effective model [Eq. (1)]. Using a dimensionless parametrization with $t_0 = 1$, the band evolution shown in Fig. 1 corresponds to $t_{am} = 0.5, 1.0, 1.5$, representing the closed-Fermi-surface, critical, and open-Fermi-surface regimes, respectively, and the DFT bands of KV$_2$Se$_2$O are consistent with the $t_{am} \gtrsim t_0$. In particular, one spin channel remains dispersive along one in-plane direction while becoming nearly flat along the orthogonal direction. In particular, the presence of Se atoms strongly suppresses the effective V-V hopping along one in-plane direction, whereas hopping along the perpendicular direction remains sizable. This anisotropic hybridization leads to an almost dispersionless band along one momentum axis, giving rise to a quasi-flat band and a saddle-point-like structure near the Fermi energy. This band anisotropy is further reflected in the spin-resolved DOS, shown in Fig. 2(d). While the total DOS remains spin compensated, consistent with the absence of net magnetization, a pronounced enhancement of the DOS appears near E$_F$, originating from the flat-band and saddle-point features. This behavior is qualitatively consistent with the logarithmic van Hove singularity predicted by the effective model in Eq. (2) for the open-Fermi-surface regime. Taken together, Figs. 2(a-d) demonstrate that KV$_2$Se$_2$O realizes, at the material level, the key ingredients of the low-energy *d*-wave altermagnetic model: direction-dependent hopping, open spin-resolved Fermi surfaces, and a strong DOS enhancement near the Fermi level. These features provide the microscopic foundation for the giant and nearly perfectly spin-polarized transport discussed in the following sections.

2.3 Tunneling Magnetoresistance in Lattice-Matched Heterostructures

In the following, we investigate two fundamental spintronic phenomena: tunneling magnetoresistance and spin-dependent thermoelectric transport, that are strongly enhanced by the nearly perfect spin polarization realized in altermagnetic spintronic devices constructed from KV$_2$Se$_2$O -based heterostructures.

To transform the crystal spin filtering of KV$_2$Se$_2$O into a functional device, we construct a fully epitaxial magnetic tunnel junction (MTJ) in which a Bi$_2$O$_2$Se barrier is sandwiched between two KV$_2$Se$_2$O electrodes in Fig. S2. Owing to the small lattice mismatch between Bi$_2$O$_2$Se and KV$_2$Se$_2$O, Bi$_2$O$_2$Se is chosen as the tunnel barrier, which preserves the intrinsic *d*-wave spin texture at the interface without introducing significant momentum scattering. Here, the parallel (PC) and antiparallel (APC) configurations refer to the relative orientation of the alternating magnetic sublattice spin textures projected along the transport direction, rather than a net magnetization alignment. In both configurations, the net magnetization remains zero. The detailed device geometry and the local spin-resolved electron density are shown in Fig. S3. Fig. 3 presents the spin-dependent transport characteristics by directly linking the microscopic energy-resolved state, described by the projected local density of states (PLDOS) [Figs. 3(a)-(d)], to the macroscopic momentum-resolved transmission [(Figs. 3(e)-(h)]. In the PC configuration, the junction exhibits a half-metallic-like transport behavior. For the spin-down channel, the PLDOS [Fig. 3(b)] reveals a continuous electronic-state bridge connecting the source and drain across the $Bi_2O_2Se$ barrier. This transportation channel connectivity gives rise to a robust transmission peak near the Fermi energy, as indicated by the red curve, which directly corresponds to a high-transparency window (bright region) in momentum space [Fig. 3(f)]. In contrast, the spin-up channel [Fig. 3(a)] is strictly blocked due to a symmetry-enforced spectral gap along the transport direction, resulting in a vanishing transmission coefficient [T(E)] and a dark momentum-resolved map [Fig. 3(e)]. As a consequence, the tunnel current becomes highly spin polarized, with a spin polarization reaching approximately 92.5% in the absence of any external field. This performance significantly surpasses that of the benchmark altermagnet $RuO_2$[22], and rivals the efficiency of conventional ferromagnetic spin injectors[57-59]. In the APC configuration, the crystal spin filter of the drain electrode is opposite to the spin-polarized current injected from the source. As shown in Figs. 3(c) and 3(d), although transport states may exist in one electrode, they lack matching counterparts in the other electrode, leading to rapid decay of the wave function at the interface. As a result, the total transmission for both spin channels is strongly suppressed, with the red curves remaining close to zero, and the momentum-resolved spectra [Figs. 3(g) and (h)] show no propagating channels. The sharp contrast between the conducting PC state and the insulating APC state gives rise to a giant tunneling magnetoresistance (TMR) ratio of approximately $2.24 \times 10^3$ % in the absence of any external magnetic field, which is more than an order of magnitude larger than the previously reported TMR of about 93% in $RuO_2$-based devices.[22]. These results establish the $KV_2Se_2O$ /$Bi_2O_2Se$ heterostructure as a promising platform for field-free spintronic logic applications.

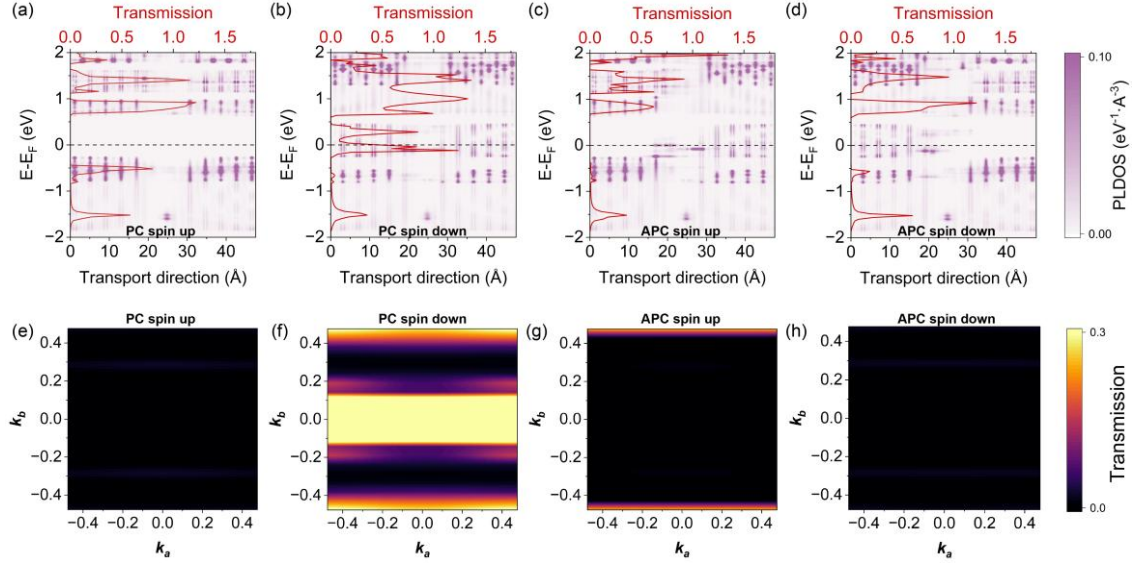

FIG. 3. Ballistic spin transport and tunneling magnetoresistance effect in a KV$_2$Se$_2$O/Bi$_2$O$_2$Se/KV$_2$Se$_2$O junction. (a–d) Energy-resolved transport mechanism. The colormaps display the Projected Local Density of States (PLDOS) along the transport direction ($k_z$), while the overlaid red curves represent the energy-dependent transmission coefficient T(E). The dashed grey line indicates the Fermi level (E$_F$). (a, b) Parallel Configuration (PC): An effective half-metallic state is realized. The spin-down channel (b) shows a continuous conducting pathway and a high transmission peak at E$_F$, whereas the spin-up channel (a) is strictly gapped with vanishing transmission. (c, d) Anti-Parallel Configuration (APC): Transport is blocked for both spins. The mismatch between the source and drain band structures leads to interface-localized states and negligible transmission coefficients (flat red curves), explaining the high resistance state. (e–h) Momentum-resolved transmission spectra at E$_F$ corresponding to the cases in (a–d). The bright yellow region in (f) confirms the highly transparent spin-down channel in the PC state, while the dark regions in (e, g, h) indicate the effective spin blockade driven by spin segregation.

## 2.4 Giant Spin-Dependent Thermoelectricity

Benefiting from the high-temperature altermagnetic nature of KV$_2$Se$_2$O, which stabilizes spin order under thermal gradients, we can explore thermally generated spin currents in KV$_2$Se$_2$O/Bi$_2$O$_2$Se/KV$_2$Se$_2$O heterostructures. Here, we explore the potential of KV$_2$Se$_2$O/Bi$_2$O$_2$Se heterostructures for spin caloritronic by investigating thermally generated spin currents. The thermal transport properties were evaluated under two distinct regimes: (i) fixed temperature difference (ΔT) with varying base left electrode temperature (T$_L$), and (ii) fixed base left electrode temperature near room temperature with varying ΔT. Fig. 4(a) illustrates the thermally induced currents under a fixed ΔT. In the parallel configuration (PC), the device operates as a highly efficient thermal spin generator. The spin-down current (red) rises significantly with temperature, driven by the Seebeck effect in the metallic minority channel. Conversely, the spin-up current (blue) remains suppressed which blocks thermally excited carriers just as effectively as electrically injected ones. This asymmetry results in a colossal thermal tunneling magnetoresistance, exceeding 750% at low temperatures and remaining substantial even at 300 K [Fig. 4(b)]. The robustness of this effect for ambient applications is demonstrated in Figs. 4(c) and 4(d), where the $T_L$ is fixed near room temperature ($T_L$ in [290, 300, 310] K). A nearly linear dependence of the spin-polarized current on $\Delta T$ is observed in the PC state, confirming the diffusive nature of the thermal transport. Most notably, the spin filtration efficiency remains remarkably high (>80%) across the entire temperature range tested [Fig. 4(d)], which indicates that the crystallographic spin segregation mechanism is energetically robust against thermal broadening, positioning $KV_2Se_2O$ as a promising candidate for zero-bias spin logic and waste-heat spin harvesting devices.

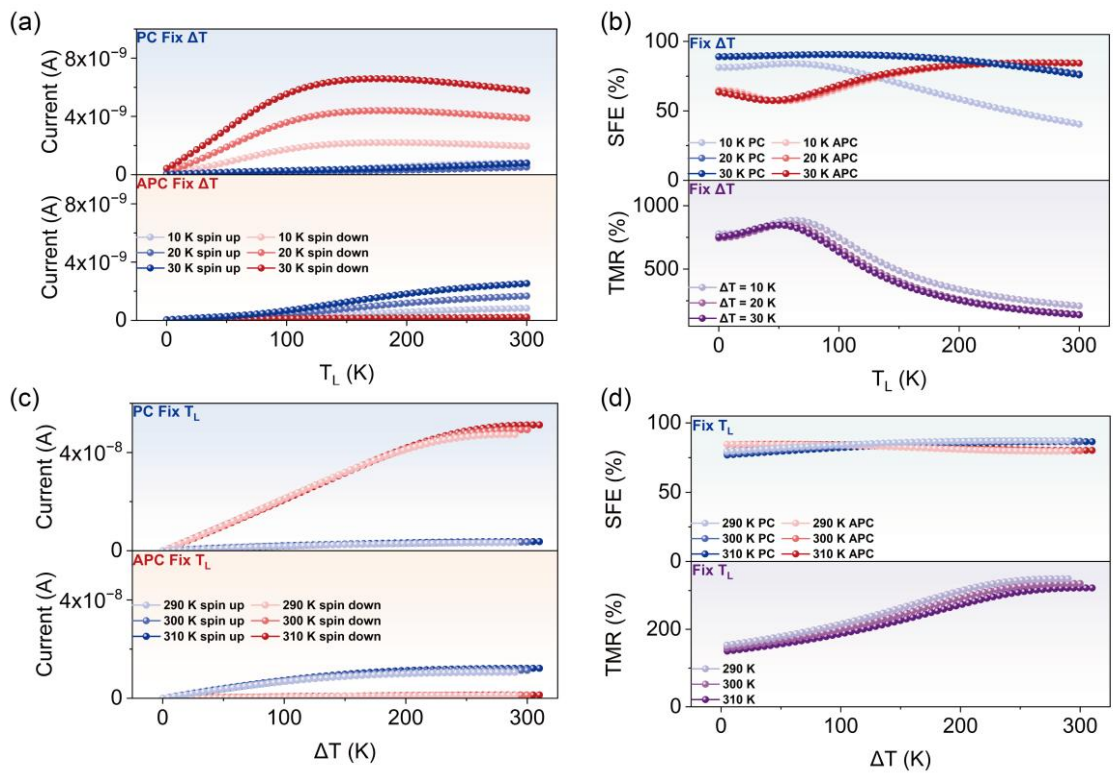

**FIG. 4. Thermally driven spin currents and magnetoresistance.** Spin-dependent thermoelectric transport properties of the $KV_2Se_2O/Bi_2O_2Se/KV_2Se_2O$ junction. (a, b) Fixed Temperature Difference ($\Delta T$). (a) Thermally generated current as a function of the left electrode temperature ($T_L$) for $\Delta T$ = 10, 20, 30 K. In the PC state, a pure spin-down current (red) is generated, while the spin-up current (blue) is blocked. (b) The corresponding spin filtration efficiency (SFE) and thermal TMR ratio. The TMR remains robust up to room temperature. (c, d) Fixed left electrode temperature ($T_L$). (c) Thermal current as a function of $\Delta T$ with $T_L$ fixed near room temperature (290–310 K). (d) The device maintains a near-unity SFE (>80%) and a rising TMR with increasing thermal gradient, confirming that the crystal orientation filtering mechanism is stable against thermal fluctuations at room temperature.

# 3 DISCUSSION

Our investigation establishes a unified framework for realizing high-performance spintronics in compensated magnetic systems, governed by the principle of crystallographic spin segregation. By bridging the symmetry-resolved toy model (Fig. 1) with specific material realizations (Figs. 2-4), we demonstrate that robust spin filtering can be rigorously enforced by magnetic crystal symmetries that allow a *d*-wave spin splitting, providing a deterministic route to generate pure spin currents without net magnetization.

Our approach is conceptually based on the distinction between direction-selective spin transport and symmetry-compensated spin splitting. While previous studies have classified altermagnets primarily based on global Fermi surface topology, our work highlights that functional spin selectivity is fundamentally dictated by local momentum-space connectivity. In $KV_2Se_2O$, the characteristic *d*-wave symmetry imposes a strict orthogonal segregation of spin channels along the principal axes. Consequently, ballistic transport along the Γ-X (or Γ-Y) direction is populated exclusively by a single spin species. This crystallographic locking effectively utilizes the anisotropic band structure to "switch on" high spin polarization simply by aligning the device geometry with the symmetry-broken crystal axes. The translation of this intrinsic quantization into macroscopic device performance is validated by the giant tunneling magnetoresistance observed in the all-epitaxial $KV_2Se_2O/Bi_2O_2Se$ heterostructure. The calculated TMR ratio of $2.24 \times 10^3$ % represents an order-of-magnitude improvement over previously reported values in $RuO_2$-based devices.

Crucially, our results extend the utility of altermagnets into the domain of spin caloritronic. The persistence of a high spin filtration efficiency (SFE> 80%) and substantial thermal TMR at room temperature indicates that the energy scale of the symmetry-protected spin splitting significantly exceeds the thermal broadening. This resilience is a direct consequence of the large spectral gaps identified in the PLDOS analysis, which effectively block thermally excited carriers in the forbidden channel. The ability to harvest pure spin currents from waste heat without external magnetic fields positions $KV_2Se_2O$ as a versatile platform for energy-efficient, zero-bias spin logic.

We emphasize that the orientation-dependent filtering mechanism identified here is generic to the *d*-wave symmetry class. By exploiting the synergy between altermagnetic ordering and epitaxial engineering, our work paves the way for a new generation of scalable, field-free, and room-temperature spintronic devices that combine the ultrafast dynamics of antiferromagnets with the high readout fidelity of ferromagnets.

## DATA AVAILABILITY

The data supporting the findings of this study are available from the corresponding authors upon reasonable request.

## ACKNOWLEDGEMENTS

This work is supported by the Shenzhen Science and Technology Program (Grant No. JCYJ20250604174400001), the Guangdong Basic and Applied Basic Research Foundation (Grant No. 2023A1515011852), and the Guangdong Science and Technology Innovation Strategy Special Fund (Grant No. pdjh2025bk012). P.-H. F. acknowledges the financial support from the Fondazione Cariplo under the grant 2023-2594. The authors would like to thank National Supercomputer Center in Guangzhou for providing high performance computational resources.

# Supplemental Material of "Intrinsic Spin Filter Effect in a *d*-wave altermagnet KV$_2$Se$_2$O with Open Fermi Surface"

Bin Liu,[1] Pei-Hao Fu,[1,2,#] Yu-Xuan Sun,[1] Xiao-Lin Zhang,[3] Si-Cong Zhu,[4] Xiang-Long Yu,[1,*] Hua Wu,[5,†] Yuan-Zhi Shao [1,6,‡]

[1] School of Science, Sun Yat-sen University, Shenzhen 518107, China
[2] Department of Physics and Astronomy, University of Florence, I-50019 Sesto Fiorentino, Italy
[3] Key Laboratory of Artificial Micro- and Nano-structures of Ministry of Education, School of Physics and Technology, Wuhan University, Wuhan 430072, China
[4] The State Key Laboratory for Refractories and Metallurgy, Wuhan University of Science and Technology, Wuhan 430081, China
[5] Laboratory for Computational Physical Sciences (MOE), State Key Laboratory of Surface Physics, and Department of Physics, Fudan University, Shanghai 200433, China
[6] Guangdong Provincial Key Laboratory of Magnetoelectric Physics and Devices, Sun Yat-sen University, Guangzhou 510275, China

[#] *Corresponding author: phy.phfu@gmail.com*
[*] *Corresponding author: yuxlong6@mail.sysu.edu.cn*
[†] *Corresponding author: wuh@fudan.edu.cn*
[‡] *Corresponding author: stssyz@mail.sysu.edu.cn*

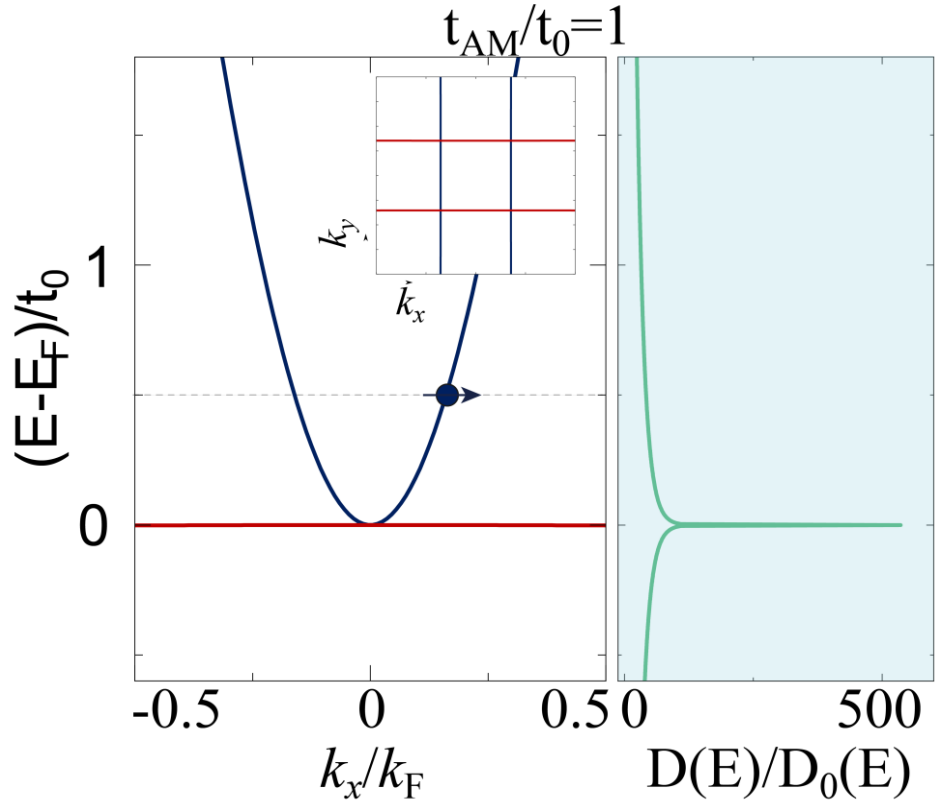

Fig. S1. Energy dispersion of the low-energy effective d-wave altermagnetic model [Eq. (1)] along $k_x$ with $k_y = 0$, together with the corresponding density of states [Eq. (2)] as a function of energy, Insets illustrate the Fermi-surface geometry determined by the altermagnetic parameter $t_{AM}/t_0$: a critical (flat) Fermi surface with zero curvature at $t_{AM}/t_0 = 1$

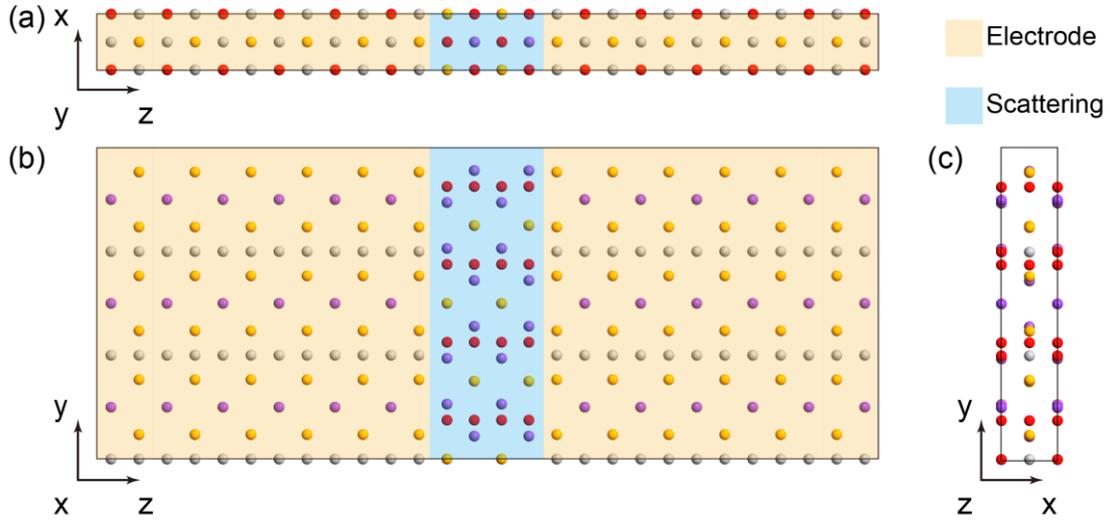

**FIG. S2.** Atomic geometry of the all-epitaxial magnetic tunnel junction. Schematic illustration of the $KV_2Se_2O/Bi_2O_2Se/KV_2Se_2O$ heterostructure used for quantum transport calculations. (a) Top view: The device consists of a central $Bi_2O_2Se$ tunnel barrier (scattering region, blue background) sandwiched between two semi-infinite $KV_2Se_2O$ electrodes (beige background). Transport is defined along the z-axis. (b) Front view: Illustrating the periodic arrangement perpendicular to the transport direction. (c) Side view: The atomic stacking viewed along the transport axis, highlighting the epitaxial registry at the interface.

Fig. S2 displays the spin-resolved local density of states (LDOS) projected onto the scattering region of the magnetic tunnel junction (MTJ). The heterostructure is constructed using $Bi_2O_2Se$ as the tunnel barrier, selected for its excellent lattice matching and chemical compatibility with the $KV_2Se_2O$ electrodes, which minimizes interface scattering and preserves the intrinsic d-wave spin texture. In the parallel configuration (PC), a striking dichotomy in wavefunction penetration is observed. The spin-down channel [Fig. S2(b)] exhibits continuous, delocalized states extending from the source electrode, through the $Bi_2O_2Se$ barrier, and into the drain, confirming the high-transparency state predicted by the orientation-dependent spin filter model. In contrast, the spin-up channel [Fig. S2(a)] shows strong localization within the electrodes with negligible spectral weight inside the barrier region, indicative of the symmetry-forbidden transport gap along the stacking direction. The blocking mechanism is further visualized in the anti-parallel configuration (APC). Here, the mismatch between the spin-projected pass-bands of the two electrodes leads to a suppression of wavefunction overlap. As shown in Supplementary Figs. S2(c, d), electronic states are confined to the injection side of the junction, decaying rapidly within the barrier without coupling to the collector. This spatial segregation of states in the APC limit, contrasted with the robust tunneling channel in the PC spin-down limit, visually corroborates the giant magnetoresistance effect arising from the crystallographic spin filtering.

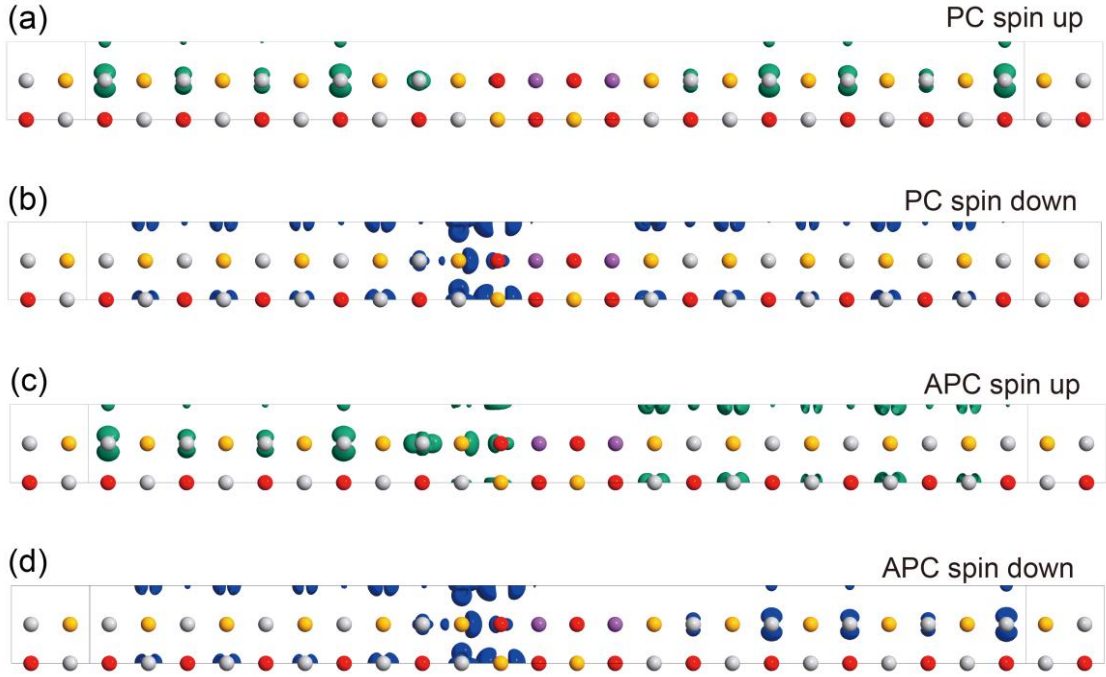

FIG. S3. Real-space projection of spin-dependent transport states. Spatial distribution of the spin-resolved Local Density of States (LDOS) at the Fermi energy for a $KV_2Se_2O/Bi_2O_2Se/KV_2Se_2O$ magnetic tunnel junction. (a, b) Parallel Configuration (PC): The spin-up channel (green) is strongly attenuated due to the absence of bulk transport states, whereas the spin-down channel (blue) demonstrates significant wavefunction penetration across the $Bi_2O_2Se$ barrier, corresponding to the high-conductance state. (c, d) Anti-Parallel Configuration (APC): Both spin-up (c) and spin-down (d) states exhibit spin-selective mismatch between the source and drain electrodes.